\documentclass[preprint,aps]{revtex4-1}
\pdfoutput=1

\usepackage{amsmath,amssymb,amsfonts,dcolumn,color,graphicx,graphics,latexsym,placeins,epsfig}
\usepackage[toc,page]{appendix}
\usepackage{comment}
\usepackage{epsfig}
\usepackage{bm}
\usepackage{slashed}
\usepackage{latexsym}
\usepackage{natbib}
\usepackage{url}
\usepackage{dcolumn}
\usepackage{color}
\usepackage{amsfonts,amssymb,amsmath}
\usepackage{graphicx,epsfig}
\usepackage{psfrag}
\usepackage{subfigure}
\usepackage{tabularx}
\usepackage{hyperref}
\hypersetup{colorlinks=true}

\newcommand{\be}{\begin{equation}}
\newcommand{\ee}{\end{equation}}
\newcommand{\ba}{\begin{eqnarray}}
\newcommand{\ea}{\end{eqnarray}}

\begin{document}

\title{Limits on dark matter, ultralight scalars, and cosmic neutrinos with gyroscope spin and precision clocks}
\author{Sara Rufrano Aliberti$^{1,2}$\footnote{s.rufranoaliberti@ssmeridionale.it}}
\author{Gaetano Lambiase$^{3,4}$\footnote{lambiase@sa.infn.it}}
\author{Tanmay Kumar Poddar$^{4}$\footnote{poddar@sa.infn.it}}
\affiliation{${}^{1}$ Scuola Superiore Meridionale, Largo S. Marcellino 10, I-80138, Napoli, Italy}
\affiliation{${}^{2}$ Istituto Nazionale di Fisica Nucleare, Sezione di Napoli, Napoli, Italy}
\affiliation{${}^{3}$ Dipartimento di Fisica ‘‘E.R. Caianiello’’, Università di Salerno, I-84084 Fisciano (SA), Italy}
\affiliation{${}^{4}$ INFN, Gruppo collegato di Salerno, I-84084 Fisciano (SA), Italy}

\begin{abstract}
Dark matter (DM) within the solar system induces deviations in the geodetic drift of gyroscope spin due to its gravitational interaction. Assuming a constant DM density as a minimal scenario, we constrain DM overdensity within the Gravity Probe B (GP-B) orbit and project limits for Earth's and Neptune's orbits around the Sun. The presence of electrons in gravitating sources and test objects introduces a scalar-mediated Yukawa potential, which can be probed using terrestrial and space--based precision clocks. We derive projected DM overdensity $(\eta)$ limits from Sagnac time measurements using onboard satellite clocks, highlighting their dependence on the source mass and orbital radius. The strongest limit, $\eta\lesssim 4.45\times 10^3$, is achieved at Neptune's orbit ($\sim 30~\mathrm{AU}$), exceeding existing constraints. Correspondingly, the cosmic neutrino overdensity is bounded as $\xi\lesssim 5.34\times 10^{10}$, surpassing results from KATRIN and cosmic ray studies. The best limit on electrophilic scalar coupling is $g\lesssim 7.09\times 10^{-24}$ for scalar mass $m_\varphi\lesssim 1.32\times 10^{-18}~\mathrm{eV}$ competitive with existing fifth-force bounds. These precision measurements offer a robust framework for testing gravity at solar system scales and probing DM in scenarios inaccessible to direct detection experiments.
\end{abstract}
  
\maketitle
\section{Introduction}\label{sec1}
Einstein's theory of General Relativity (GR) provides an exceptionally accurate description of gravity, passing all experimental tests and consistently aligning with observations at an unprecedented level of precision. Among its key predictions are the geodetic effect \cite{deSitter:1916zz} and frame-dragging \cite{Lense1918,Mashhoon:1984fj}, both of which have been experimentally verified, notably by the Gravity Probe-B (GP-B) mission \cite{Buchman:2000quk,Everitt:2015qri}.

The geodetic effect (de Sitter precession) and frame-dragging (Lense-Thirring effect) describe spacetime curvature and twisting caused by a massive object's mass and rotation, respectively \cite{Schiff:1960gh}. The GP-B experiment, launched in 2004, used ultra-precise gyroscopes in Earth's orbit ($\sim 640~\mathrm{km}$ altitude) to measure these effects \cite{gpb}. It observed a geodetic drift of $-6601.8\pm 18.3~\mathrm{mas/yr}$  and a frame-dragging drift of $-37.2\pm 7.2~\mathrm{mas/yr}$ closely aligning with GR predictions of $-6606.1~\mathrm{mas/yr}$ and $-39.2~\mathrm{mas/yr}$ \cite{Everitt:2011hp}.

Beyond GP-B, several experiments, such as Lunar Laser Ranging (LLR) \cite{Bertotti:1987zz,Williams:1995nq}, LARES-2 \cite{Capozziello:2014mea,Ciufolini:2023czv}, LAGEOS-2 \cite{Iorio:2002rm,Ciufolini:2004rq,Iorio:2008vm}, VLBI \cite{Klopotek2020,SCHUH201268}, GINGER \cite{Altucci:2022rxr,DiVirgilio:2023nrc}, GRACE  \cite{doi:10.1126/science.1099192}, and GRACE-FO \cite{Abich:2019cci}, aim to refine measurements of geodetic and frame-dragging effects with high precision, further validating GR. These efforts have also been used to constrain alternative gravity theories and Beyond Standard Model (BSM) scenarios through gyroscope precession \cite{Lambiase:2013dai, Radicella:2014jwa, Capozziello:2014mea, Poddar:2021ose, Benisty:2023dkn}.

Precision experiments provide powerful tools to constrain Dark Matter (DM), which constitutes $25\%$ of the universe's energy density \cite{Planck:2018vyg}. DM is inferred through gravitational interactions across scales, from galactic rotation curves \cite{Rubin:1970zza,Rubin:1978kmz,Rubin:1980zd,Persic:1995ru,Corbelli:1999af} to cosmological observations like the Cosmic Microwave Background (CMB) \cite{WMAP:2008lyn}. The Bullet Cluster offers strong evidence for DM's particle nature \cite{Clowe:2003tk,Markevitch:2003at}, and DM plays a key role in cosmic structure formation \cite{Blumenthal:1984bp}.

DM particle masses range from $10^{-22}~\mathrm{eV}$ to several solar masses \cite{Cirelli:2024ssz}, with a local density near the Sun of approximately $\bar{\rho}\simeq 0.4~\mathrm{GeV/cm^3}$ \cite{Catena:2009mf,Weber:2009pt,Bovy:2012tw,Nesti:2012zp,Read:2014qva,deSalas:2019pee}. This density varies due to quantum fluctuations during inflation and gravitational instabilities. While DM's non-interaction with photons makes direct detection difficult, gravitational lensing provides indirect evidence. BSM scenarios predict specific DM structures, such as boson stars \cite{Colpi:1986ye,Tkachev:1991ka,Kolb:1993zz,Chen:2020cef,Dmitriev:2021utv,Visinelli:2021uve}, axion miniclusters \cite{Kolb:1993zz,Kolb:1995bu,Tkachev:2014dpa,Dokuchaev:2017psd,Eggemeier:2019jsu,Eggemeier:2019khm,Kirkpatrick:2020fwd,Kavanagh:2020gcy,Xiao:2021nkb}, and others \cite{Budker:2019zka,Lasenby:2020goo,VanTilburg:2020jvl,Wu:2022wzw} offering further avenues for exploration.

This paper investigates DM overdensities using only gravitational interactions, characterized by $\eta=\rho/\bar{\rho}$. Existing constraints on $\eta$ range from $10^4$ to $10^8$, depending on the scale, with perihelion precession and asteroid tracking providing notable bounds \cite{Gron:1995rn, Tsai:2022jnv}. DM overdensities near Black Holes (BHs) can also be constrained via Gravitational Wave (GW) observations \cite{Kavanagh:2020cfn, Coogan:2021uqv}.

DM within the solar system could influence gyroscope spin precession, altering geodetic and frame-dragging effects. High-precision measurements of these effects offer a way to constrain DM density. Additionally, a DM-induced force, distinct from gravitational forces, may arise in Schwarzschild spacetime. Its radial dependence could deviate from the inverse-square law, providing opportunities to detect such forces via fifth-force and equivalence principle tests \cite{Fischbach:1996eq,Adelberger:2003zx,Berge:2017ovy,Hees:2018fpg}.

Earth and space--based precision clocks offer a unique method to search for non-gravitational potentials and DM-induced effects. Variations in atomic or nuclear transition frequencies, influenced by a spatially dependent DM potential, can be analyzed using atomic and nuclear clocks \cite{Barrow:1999qk, Dzuba:2024pri,Caputo:2024doz}. Additionally, the ellipticity of a planetary orbit and the Sagnac time measurements with the onboard precision clocks in a DM-filled background can reveal deviations caused by DM. The Sagnac effect, arising from differential light travel times around a rotating source, provides a measurable shift in interference fringes due to DM's influence on spacetime \cite{Sagnac, POST:1967qwl, Souza:2024ltj}.

Geodetic and frame-dragging measurements also enable exploration of long-range Yukawa potentials mediated by ultralight scalar field. In the non-relativistic limit, if an ultralight scalar field couples with electrons, it would generate a long-range Yukawa potential \cite{Moody:1984ba}. Ultralight scalar DM can arise from mechanisms such as the misalignment mechanism or the Stueckelberg mechanism \cite{Nelson:2011sf,Hui:2016ltb}, with its small mass explained by frameworks like clockwork models or D-term inflation \cite{Fayet:1974jb,Fayet:2017pdp,Joshipura:2020ibd}. These ultralight candidates interact weakly with SM particles, and their coupling strengths are tightly constrained by a range of experiments and observations \cite{AxionLimits}. These scalar particles in the non-relativistic limit can induce fifth forces that deviate from the $1/r^2$ gravitational law. The coupling strength of scalars can be constrained through their interactions with electrons in the source and test object. Accurate measurements of gyroscope spin dynamics and precision clocks further enhance the ability to probe these forces, providing insights into DM and ultralight scalars.

The spin motion of gyroscopes offers a complementary method to probe the Cosmic Neutrino Background (C$\nu$B), a relic from the early universe with neutrino energies of $10^{-4}~\mathrm{eV}$ to $10^{-6}~\mathrm{eV}$ \cite{Follin:2015hya}. These neutrinos, decoupled since the universe was one second old, provide insights into the pre-CMB epoch. Two neutrino species are non-relativistic today, potentially forming overdense regions akin to DM. The cosmic neutrino overdensity can be parametrized as $\xi=\rho_\nu/\bar{\rho}_\nu$, where $\bar{\rho}_\nu=m_\nu \bar{n}_\nu$ with $m_\nu$ the neutrino mass and total cosmic neutrino number density $\bar{n}_\nu=336/\mathrm{cm^3}$ (considering all flavours of neutrinos and anti-neutrinos), predicted by SM. Ground and space-based precision clocks can also put limits on cosmic neutrino overdensity.

Experiments like PTOLEMY aim to detect the C$\nu$B via inverse tritium beta decay \cite{Betts:2013uya}. Constraints on C$\nu$B overdensities arise from cosmic ray studies \cite{Ciscar-Monsalvatje:2024tvm}, the CMB \cite{Follin:2015hya}, KATRIN \cite{KATRIN:2022kkv}, perihelion precession of celestial objects \cite{Tsai:2022jnv}, and others \cite{Bauer:2022lri}. Such overdensities could affect gyroscope motion, offering a novel way to test GR and constrain C$\nu$B properties.

The paper is structured as follows: Section \ref{sec2} derives the geodetic precession rate in Schwarzschild spacetime with constant DM density. Section \ref{sec3} extends this to include Yukawa potentials mediated by ultralight scalar. Section \ref{sec4} explores forces induced by DM and long-range interactions. Section \ref{sec5} examines DM and Yukawa effects on atomic and nuclear transition frequency ratios. Section \ref{sec6} derives the Sagnac time delay due to DM density. Section \ref{sec8} discusses the results obtained in this study. Section \ref{addneu} discusses limits on C$\nu$B overdensity, and Section \ref{sec9} concludes with a summary and implications.

Natural units are used throughout, with $G=1$ unless stated otherwise.

\section{Geodetic precession in presence of dark matter filled Schwarzschild background}\label{sec2}
In the following, we examine the geodetic precession of a gyroscope's spin in the Schwarzschild spacetime around a gravitating source. The analysis accounts for the effect of a surrounding constant DM background, providing a comprehensive understanding of its effects. The static spherically symmetric Schwarzschild metric in presence of constant DM density $(\rho_0)$ is \cite{Gron:1995rn}
\begin{equation}
ds^2=-\Big(1-\frac{2M}{r}+\frac{4\pi}{3}\rho_0 r^2\Big)dt^{2}+\frac{dr^2}{1-\frac{2M}{r}-\frac{8\pi}{3}\rho_0r^2}dr^2+r^2d\theta^2+r^2\sin^2\theta d\phi^2,
\label{eq:1}
\end{equation}
where $M$ denotes the mass of the gravitating source and $t$, $r$, $\theta$, $\phi$ denote the time, radial, polar and azimuthal coordinates. The above metric is valid outside the source and for $r\ll(M/\rho_o)^{1/3}$. We consider that the DM is non-relativistic matter and its density is much larger than its pressure.

There are two conserved quantities associated with the two killing vectors $\xi_t=(1,0,0,0)$ and $\xi_\phi=(0,0,0,1)$. Therefore, the conserved quantities are
\begin{equation}
E=\xi^\mu_t u_\mu=g_{\mu\nu}\xi^\mu_t u^\nu=g_{00}u^0\xi^0_t=\Big(1-\frac{2M}{r}+\frac{4\pi}{3}\rho_0r^2\Big)\dot{t},
\label{eq:a}
\end{equation}
and
\begin{equation}
L=\xi^\mu_\phi u_\mu=g_{\mu\nu}\xi^\mu_\phi u^\nu=g_{33}u^3\xi^3_\phi=r^2\dot{\phi},
\label{eq:a1k}
\end{equation}
where $E$ and $L$ are the energy and angular momentum of the system per unit mass and $u^\mu=\frac{dx^\mu}{d\tau}=(u^0,u^1, u^2, u^3)$ denotes the four velocity. The overdot represents the derivative with respect to the proper time $(\tau)$. We consider the motion of the gyroscope around the source is in the equatorial plane and hence, $\theta=\pi/2$. The energy equation for the gyroscope orbiting around the source in the spacetime described by Eq. \ref{eq:1} is 
\begin{equation}
\Big(1-\frac{2M}{r}+\frac{4\pi}{3}\rho_0r^2\Big)\dot{t}^2-\Big(1-\frac{2M}{r}-\frac{8\pi}{3}\rho_0r^2\Big)^{-1}\dot{r}^2-r^2\dot{\phi}^2=1.
\label{eq:b}
\end{equation}
Using Eqs. \ref{eq:a} and \ref{eq:a1k}, we can write Eq. \ref{eq:b} as

\begin{equation}
\frac{E^2-1}{2}=\frac{\dot{r}^2}{2}+\frac{L^2}{2r^2}-\frac{ML^2}{r^3}-\frac{M}{r}+\frac{2\pi}{3}\rho_0r^2-\frac{4\pi\rho_0L^2}{3},
\label{eq:c}
\end{equation}
where higher order terms are neglected. Thus, the effective potential of the system becomes
\begin{equation}
V_{\mathrm{eff}}(r)=\frac{L^2}{2r^2}-\frac{ML^2}{r^3}-\frac{M}{r}+\frac{2\pi}{3}\rho_0r^2-\frac{4\pi\rho_0L^2}{3}.
\label{eq:d}
\end{equation}
For a circular orbit of radius $R$, $\frac{dV_{\mathrm{eff}}}{dr}=0$ at $r=R$ and we obtain the expression of $L$ from Eq. \ref{eq:d} as
\begin{equation}
L^2=\frac{MR^2}{R-3M}\Big(1+\frac{4\pi}{3}\frac{\rho_0}{M}R^3\Big).
\label{eq:e}
\end{equation}
At the minimum of the potential, $\dot{r}=0$, and we can write the energy  of the system for the gyroscope orbit of radius $R$ from Eq. \ref{eq:c} and using Eq. \ref{eq:e} as
\begin{equation}
E^2=\Big(1-\frac{M}{R}+\frac{8\pi}{3}\rho_0R^2\Big).
\label{eq:f}
\end{equation}
The angular velocity of the gyroscope is defined as
\begin{equation}
 \Omega=\frac{d\phi}{dt}=\frac{d\phi/d\lambda}{dt/d\lambda}=\frac{L}{ER^2}\Big(1-\frac{2M}{R}+\frac{4\pi}{3}\rho_0 R^2 \Big).
 \label{eq:a1}
\end{equation}
Using Eqs. \ref{eq:e} and \ref{eq:f}, we obtain the expression for angular velocity from Eq. \ref{eq:a1} as
\begin{equation}
\Omega=\frac{\sqrt{M}}{R^{3/2}}\Big(1+\frac{2\pi}{3}\frac{\rho_0R^3}{M}\Big), 
\label{eq:a2}
\end{equation}
where in the limit $\rho_0\rightarrow 0$, we get the standard angular velocity for a Keplerian orbit in GR.

To study the dynamics of the gyroscope spin, we define its spin vector as $S^\mu=(S^{(0)}, S^{(1)}, S^{(2)}, S^{(3)})$. In the rest frame of the gyroscope, $S^\mu=(0,\mathbf{S})$ and $u^\mu=(u^0,\mathbf{0})$ and hence, $S^\mu u_\mu=0$. In fact, we can define $u^\mu$ and $S^\mu$ in such a way that in any frame $g_{\mu\nu}u^\mu S^\nu=0$. Since, there is no radial movement of the gyroscope as it is in the stable circular orbit and the motion is confined in the plane, we must have $u^1=u^2=0$. Therefore, $u^\mu=(u^0,0,0,u^3)=u^0(1,0,0,\Omega)$, where $\Omega=u^3/u^0=d\phi/dt$. 

As the gyroscope is a massive object and it is orbiting around the source, we can write
\begin{equation}
\Big(1-\frac{2M}{R}+\frac{4\pi}{3}\rho_0 R^2\Big){u^0}^2-R^2{u^0}^2\Omega^2=1,
\label{eq:4}
\end{equation}
which follows from $g_{\mu\nu}u^\mu u^\nu=-1$. Thus, the solution of $u^0$ becomes
\begin{equation}
u^0=\frac{1}{\sqrt{1-\frac{2M}{R}+\frac{4\pi}{3}\rho_0 R^2-R^2\Omega^2}}.
\label{eq:5}
\end{equation}
From $g_{\mu\nu}u^\mu S^\nu=0$, we can write
\begin{equation}
\Big(1-\frac{2M}{R}+\frac{4\pi}{3}\rho_0R^2\Big)u^0S^{(0)}-R^2u^3S^{(3)}=0.
\label{eq:6}
\end{equation}
Using $u^3=\Omega u^0$, we can write Eq. \ref{eq:6} as
\begin{equation}
S^{(0)}=\frac{R^2\Omega S^{(3)}}{\Big(1-\frac{2M}{R}+\frac{4\pi}{3}\rho_0R^2\Big)}.
\label{eq:7}
\end{equation}
Also, the spin vector $S^\mu$ follows the parallel transport and we can write
\begin{equation}
\frac{dS^\mu}{d\tau}+\Gamma^{\mu}_{\alpha\beta}u^\alpha S^\beta=0.
\label{eq:3}
\end{equation}
Since, there are only seven non zero Christoffel symbols (see Appendix \ref{app1}), we can write the gyroscope equations for all the spin components as 
\begin{equation}
\begin{split}
\frac{dS^{(0)}}{d\tau}+\Gamma^0_{01}u^0S^{(1)}=0,\\
\frac{dS^{(1)}}{d\tau}+\Gamma^1_{00}u^0S^{(0)}+\Gamma^1_{33}u^3S^{(3)}=0,\\
\frac{dS^{(2)}}{d\tau}=0,\\
\frac{dS^{(3)}}{d\tau}+\Gamma^3_{31}u^3S^{(1)}=0.
\end{split}
\label{eq:es1}
\end{equation}
Using Eqs. \ref{eq:es1} and \ref{eq:2} we obtain the time evolution equation for $S^{(1)}$ as
\begin{equation}
\frac{dS^{(1)}}{d\tau}+(3M-R+4\pi\rho_0R^3)\Omega u^0S^{(3)}=0.
\label{eq:8}
\end{equation}
Similarly, from Eq. \ref{eq:es1} and using Eq. \ref{eq:2} we can write the time evolution equation for $S^{(3)}$ as 
\begin{equation}
\frac{dS^{(3)}}{d\tau}=-\frac{\Omega u^{0}S^{(1)}}{R}.
\label{eq:9}
\end{equation}
Further differentiating Eq. \ref{eq:8} and using Eq. \ref{eq:9} we obtain
\begin{equation}
\frac{d^2S^{(1)}}{d\tau^2}+(R-3M-4\pi\rho_0R^3)\frac{\Omega^2{u^0}^2S^{(1)}}{R}=0.
\label{eq:10}
\end{equation}
Using Eqs. \ref{eq:a2} and \ref{eq:5}, we can write Eq. \ref{eq:10} as
\begin{equation}
\frac{d^2S^{(1)}}{dt^2}+\frac{\Omega^2_f}{{u^0}^2}S^{(1)}=0,
\label{eq:11}
\end{equation}
where we use the fact $u^0=dt/d\tau$ and 
\begin{equation}
\Omega_f=\Omega\Big(1-4\pi\rho_0R^2\Big)^\frac{1}{2}.
\label{eq:12}
\end{equation}
Therefore, the solution of Eq. \ref{eq:11} becomes
\begin{equation}
S^{(1)}(t)=S^{(1)}(0)\cos\Big(\frac{\Omega_f t}{u_0}\Big),  
\label{eq:13}
\end{equation}
where we choose the initial condition of the spin four vector as $S^\mu=(S^{(0)}(0), S^{(1)}(0), 0, 0)$ at $t=0$.

Integrating Eq. \ref{eq:9} and using Eq. \ref{eq:13} we obtain the time evolution of $S^{(3)}$ as
\begin{equation}
S^{(3)}(t)=-\frac{u^0}{R}S^{(1)}(0)\Big(1+2\pi\rho_0R^2\Big)\sin\Big(\frac{\Omega_f}{u^0}t\Big).
\label{eq:14}
\end{equation}
Integrating Eq. \ref{eq:es1} and using the initial condition, the solution of $S^{(2)}$ is $S^{(2)}(t)=0$. 

To obtain the solution of $S^{(0)}$, we use Eqs. \ref{eq:es1} and \ref{eq:2} and get
\begin{equation}
\frac{dS^{0}}{dt}+\Big(1+\frac{2M}{R}-\frac{4\pi}{3}\rho_0R^2\Big)\Big(\frac{M}{R^2}+\frac{4\pi}{3}\rho_0R\Big)S^{(1)}(0)\cos\Big(\frac{\Omega_f}{u^0}t\Big)=0.
\label{eq:15}
\end{equation}
Integrating Eq. \ref{eq:15} we obtain
\begin{equation}
S^{(0)}(t)=S^{(0)}(0)-S^{(1)}(0)\frac{u^0}{\Omega}\Big(\frac{M}{R^2}+\frac{4\pi}{3}\rho_0R\Big)\sin\Big(\frac{\Omega_f}{u^0}t\Big).
\label{eq:16}
\end{equation}
At $t=2\pi/\Omega$, i.e; time after one complete orbit, we can write Eq. \ref{eq:13} as
\begin{equation}
S^{(1)}\Big(\frac{2\pi}{\Omega}\Big)=S^{(1)}(0)\cos\Big(\frac{2\pi}{u^0}\Big(1-4\pi\rho_0 R^2\Big)^\frac{1}{2}\Big).
\label{eq:17}
\end{equation}
Therefore, the advancement of $S^{(1)}$ is
\begin{eqnarray}
\alpha&=&2\pi-\frac{2\pi}{u^0}\Big(1-4\pi\rho_0R^2\Big)^\frac{1}{2}\nonumber\\
&=& 2\pi\Big[1-\frac{1}{u^0}\Big(1-2\pi\rho_0 R^2\Big)\Big],
\label{eq:18}
\end{eqnarray}
where the expressions for $\Omega$ and $u^0$ are obtained from Eqs. \ref{eq:a2} and \ref{eq:5} respectively and we can write
\begin{equation}
\alpha=\frac{3\pi M}{R}+4\pi^2\rho_0R^2,  
\end{equation}
where in the limit, $\rho_0\rightarrow 0$, $\alpha=\frac{3\pi M}{R}$, which matches with the standard GR result. This is the expression for the geodetic precession of the gyroscope around a gravitating source in presence of DM.

\section{Geodetic precession due to the presence of a scalar mediated Yukawa potential in the Schwarzschild spacetime background}\label{sec3}
\begin{figure}[htp]
    \centering
    \includegraphics[width=12cm]{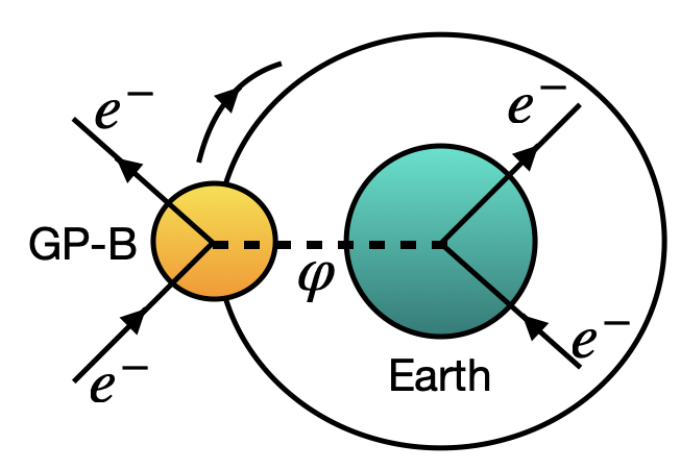}
    \caption{Schematic plot of an electrophilic scalar mediated Feynman diagram between the Earth and the GP-B}
    \label{p1}
\end{figure}
If an ultralight scalar field background is coupled with the electrons in the non-relativistic gravitating source and the gyroscope then it can give rise to a long-range Yukawa-type potential. In FIG. \ref{p1}, we show the schematic plot of an electrophilic scalar mediated Feynman diagram between the Earth and the GP-B gyroscope. The ultralight scalar may or may not contribute to the DM energy density in the universe. This interaction creates a coherent force between the electrons, resulting in a Yukawa potential whose range exceeds the distance between the source and the gyroscope. Consequently, the motion of the gyroscope around the source, under the influence of this non-gravitational scalar mediated long-range force, can be described by the following action
\begin{equation}
S = -M_{\mathrm{gy}}\int \sqrt{-g_{\mu\nu}\dot{x}^\mu \dot{x}^\nu} d\tau -g q\int \varphi \dot{t} d\tau,
\label{action}
\end{equation}
 where $g_{\mu\nu}$ denotes the metric tensor for the Schwarzschild background, the overdot stands for the derivative with respect to the proper time $\tau$, $M_{\mathrm{gy}}$ is the mass of the gyroscope, $g$ is the coupling constant which couples the classical electron number density of the gyroscope with the scalar field $\varphi$ due to the source, and $q$ is the total charge due to the presence of electrons in the gyroscope. Varying the action Eq. \ref{action}, we obtain the equation of the motion of the gyroscope as
\begin{equation}
 \Ddot{x}^\alpha+\Gamma^\alpha_{\mu\nu}\dot{x}^\mu\dot{x}^\nu=\frac{g\,q}{M_{gy}}\left(\dot{t}\,g^{\alpha\lambda}\partial_\lambda\varphi -g^{\alpha 0} \dot{x}^\mu\partial_\mu\varphi \right),
\label{geodetic+interaction}
\end{equation}
where we study the static case, and $\varphi(r)\approx \frac{gQ}{4\pi r}e^{-m_\varphi r}$ \cite{Leefer:2016xfu} is generated due to the presence of electrons in the gravitating source. Here, $Q$ is the total charge due to the presence of electrons in the source and $m_\varphi$ is the mass of the ultralight scalar. We write the line element for the Schwarzschild metric outside the source as
\begin{equation}
ds^2 = -\left( 1-\frac{2M}{r} \right)dt^2+\frac{1}{\left( 1-\frac{2M}{r}\right)}dr^2+r^2d\theta^2+r^2\sin^2\theta d\phi^2.
\end{equation}
 The temporal part of Eq. \ref{geodetic+interaction} is given as 
\begin{equation}
    \ddot{t}+\frac{2M}{r^2\left( 1-\frac{2M}{r} \right)}\dot{r}\dot{t} = \frac{gq}{M_{\mathrm{gy}}\left( 1-\frac{2M}{r}\right)}\frac{d\varphi}{dr}\dot{r},
\end{equation}
whose solution is obtained as
\begin{equation}
\dot{t} = \frac{\left( E_\varphi+\frac{g q\,\varphi}{M_{\mathrm{gy}}} \right)}{\left(1-\frac{2M}{r}\right)},
\label{tdot}
\end{equation}
where $E_\varphi$ is a constant of motion, interpreted as the total energy per unit rest mass.

Considering the motion of the gyroscope is confined in a plane, we can write the azimuthal part of Eq. \ref{geodetic+interaction} as
\begin{equation}
    \ddot{\phi} + \frac{2}{r}\dot{r}\dot{\phi} = 0,
\end{equation}
and its solution is
\begin{equation}
    \dot{\phi} = \frac{L_\varphi}{r^2},
    \label{phidot}
\end{equation}
where $L_\varphi$ is a constant of motion, denotes as the angular momentum per unit rest mass.

As the gyroscope follows timelike geodesic, $g_{\mu\nu}\,\dot{x}^\mu\dot{x}^\nu = -1$, we can write 
\begin{equation}
    \frac{\left(E_\varphi+\frac{gq\,\varphi}{M_{gy}}\right)^2-1}{2} = \frac{\dot{r}^2}{2}+\frac{L_\varphi^2}{2r^2}-\frac{M\,L_\varphi^2}{r^3}-\frac{M}{r},
\end{equation}
where we use Eqs. \ref{tdot} and \ref{phidot}. Therefore, using the expression of $\varphi(r)$ and neglecting the higher order terms, we obtain
\begin{equation}
 \frac{E_\varphi^2-1}{2} = \frac{\dot{r}^2}{2} +\frac{L_\varphi^2}{2r^2} -\frac{ML_\varphi^2}{r^3} -\frac{M}{r} -\frac{g^2qQ}{4\pi M_{gy}r}e^{-m_\varphi r}.
    \label{square_energy}
\end{equation}
Thus, the effective potential of the system can be written as
\begin{equation}
    V_{\mathrm{eff}}(r) =\frac{L_\varphi^2}{2r^2} -\frac{ML_\varphi^2}{r^3} -\frac{M}{r} -\frac{g^2qQ}{4\pi M_{\mathrm{gy}}r}e^{-m_\varphi r}.
    \label{eff}
\end{equation}
To obtain the solution of $L_\varphi$, we consider $dV_\mathrm{eff}/dr=0$, at $r=R$ for a circular orbit of radius $R$ and we get
\begin{equation}
 L_\varphi^2=\frac{R^2}{R-3M}\Big[M+\frac{g^2Qq}{4\pi M_\mathrm{gy}}e^{-m_\varphi R}(1+m_\varphi R)\Big]. 
 \label{angu}
\end{equation}
At the minimum of the potential, $\dot{r}=0$ and we can express the energy per unit mass for a circular orbit of radius $R$ from Eq. \ref{square_energy} as 
\begin{equation}
E_\varphi=1-\frac{M}{2R}-\frac{g^2 qQ}{8\pi M_{\mathrm{gy}}R}e^{-m_\varphi R}(1-m_\varphi R),
\label{en}
\end{equation}
where we use Eq. \ref{angu} and neglect the higher order terms. Therefore, we can calculate the angular velocity of the gyroscope as
\begin{equation}
\Omega=\frac{d\phi}{dt}=\sqrt{\frac{M}{R^3}}\Big[1-\frac{g^2qQ}{4\pi M_\mathrm{gy}R}e^{-m_\varphi R}+\frac{g^2qQ}{8\pi M_{\mathrm{gy}}M}e^{-m_\varphi R}(1+m_\varphi R)\Big],
\label{ang}
\end{equation}
where in the limit $g\rightarrow 0$, we recover the standard Keplerian result.

The spin vector of the gyroscope satisfies the condition $g_{\mu\nu}u^\mu S^\nu=0$. When the gyroscope is in a stable circular orbit with its motion confined to a plane, this condition simplifies, leading to the following result as
\begin{equation}
 S^{(0)}=\frac{R^2\Omega S^{(3)}}{\Big(1-
 \frac{2M}{R}\Big)}.   
 \label{mn1}
\end{equation}
The gyroscope is also a massive object which satisfies $g_{\mu\nu}u^\mu u^{\nu}=-1$, and we obtain
\begin{equation}
 u_0=\frac{1}{\sqrt{1-\frac{2M}{R}-R^2\Omega^2}}.
 \label{mn2}
\end{equation}
We can write the equation of motion of the gyroscope's spin in presence of long-range Yukawa potential as
\begin{equation}
 \frac{dS^\mu}{d\tau} + \Gamma^\mu_{\alpha\beta}\,u^\alpha S^\beta =\frac{gq}{M_{gy}}g^{\mu\lambda}\left( \partial_\lambda \varphi\delta_\nu^0 -\partial_\nu \varphi\delta_\lambda^0 \right)\, S^\nu,  
 \label{spin_eq_Yukawa}
\end{equation}
which is derived in Appendix \ref{app2}. Considering the same initial conditions as chosen in Section \ref{sec2}, we obtain the equation of motion of spin $S^{(1)}$ as
\begin{equation}
 \frac{dS^{(1)}}{d\tau}+\Omega u^0 S^{(3)}(3M-R)=\frac{gq}{M_{\mathrm{gy}}}R^2 \Omega S^{(3)}\Big(\frac{\partial \varphi}{\partial r}\Big)\Bigg|_{r=R},   
 \label{res1a}
\end{equation}
and the equation of motion for $S^{(3)}$ is
\begin{equation}
\frac{dS^{(3)}}{d\tau}=-\frac{1}{R}\Omega u^0 S^{(1)}. 
\label{res2a}
\end{equation}
Combining Eqs. \ref{res1a} and \ref{res2a}, we obtain
\begin{equation}
\frac{d^2S^{(1)}}{dt^2}+\Omega^2_f S^{(1)}=0,   
\label{res3a}
\end{equation}
where 
\begin{equation}
 \Omega^2_f=\Omega^2 \Big[1-
 \frac{3M}{R}+\frac{gqR}{u^0 M_{\mathrm{gy}}}\Big(\frac{\partial \varphi}{\partial r}\Big)\Bigg|_{r=R}\Big]. 
 \label{res4}
\end{equation}
Using the expression of $\varphi(r)$ and Eq. \ref{mn2}, the expression of Eq. \ref{res4} simplifies to 
\begin{equation}
 \Omega^2_f=\Omega^2\Big[1-
 \frac{3M}{R}-\frac{g^2qQ}{4\pi R M_{\mathrm{gy}}}e^{-m_\varphi R}(1+m_\varphi R)\Big],  
 \label{res5}
\end{equation}
where we avoid the higher order terms. Therefore, the solution of Eq. \ref{res3a} becomes
\begin{equation}
 S^{(1)}(t)=S^{(1)}(0)\cos(\Omega_f t).   
\label{res6}
\end{equation}
Similarly, we obtain the solution of Eq. \ref{res2a} using Eqs. \ref{res5} and \ref{res6} as
\begin{equation}
 S^{(3)}(t)=-\frac{S^{(1)}(0)}{R}\Big[1+\frac{3M}{2R}+\frac{g^2qQ}{8\pi RM_{\mathrm{gy}}}e^{-m_\varphi R}(1+m_\varphi R)\Big]\sin(\Omega_f t).   
 \label{res7}
\end{equation}
We also have the solution for $S^{(2)}(t)$ as
\begin{equation}
S^{(2)}(t)=0.  
\label{res8}
\end{equation}
The solution of $S^{(0)}(t)$ can be obtained from Eq. \ref{spin_eq_Yukawa} as
\begin{equation}
S^{(0)}(t)=S^{(0)}(0)-\frac{S^{(1)}(0)M\sin(\Omega_f t)}{R^2\Omega}\Big[1+\frac{g^2qQ}{4\pi M_{\mathrm{gy}}M}e^{-m_\varphi R}(1+m_\varphi R)\Big].   
\label{res9}
\end{equation}
At $t=2\pi/\Omega$, i.e; time after one complete orbit, we write Eq. \ref{res6} as 
\begin{equation}
S^{(1)}\Big(\frac{2\pi}{\Omega}\Big)=S^{(1)}(0)\cos\Big(2\pi \frac{\Omega_f}{\Omega}\Big).   
\label{res10}
\end{equation}
Therefore, the advancement of $S^{(1)}$ is
\begin{equation}
\alpha= 2\pi\Big(1-\frac{\Omega_f}{\Omega}\Big)= \frac{3\pi M}{R}+\frac{g^2qQ}{4RM_{\mathrm{gy}}}e^{-m_\varphi R}(1+m_\varphi R),
\label{res11}
\end{equation}
where in the limit $g\rightarrow 0$, the standard GR result is obtained. A similar analysis can be done with the mediation of ultralight vectors coupled with non-relativistic electrons and the results will be exactly same in the static limit \cite{KumarPoddar:2020kdz}. 
\section{Search for a new force}\label{sec4}
The contribution of DM to the potential per unit mass for the source-gyroscope system immersed in a constant DM background is given as
\begin{equation}
 V_\mathrm{DM}(r)=\frac{2\pi}{3}\rho_0 r^2, 
 \label{force1}
\end{equation}
where we use Eq. \ref{eq:d} and avoid the higher order terms. Therefore, the magnitude of the DM induced force per unit mass is obtained as 
\begin{equation}
 F_{\mathrm{DM}}(r)=\frac{4\pi}{3}\rho_0 r,
 \label{force2}
\end{equation}
where the DM force grows with distance, unlike the gravitational force, which follows an infinite-range $1/r^2$ behaviour. The DM induced force acts along the radial direction $(\hat{\mathbf{r}})$, i.e., along the line joining the source and the gyroscope.

Similarly the scalar mediated potential contribution of the source-gyroscope system is obtained from Eq. \ref{square_energy} as
\begin{equation}
V_{L}(r)=-\frac{g^2qQ}{4\pi M_{\mathrm{gy}}r}e^{-m_\varphi r},  
\label{force3}
\end{equation}
and the corresponding long-range Yukawa type force per unit mass is obtained as
\begin{equation}
F_L(r)=\frac{g^2qQ}{4\pi M_{\mathrm{gy}}r^2}(1+m_\varphi r)e^{-m_\varphi r}   
\label{force4}
\end{equation}
where the range of the force is given as $\lambda>1/m_\varphi$ which is basically the distance between the source and the gyroscope.
\section{Search for non-gravitational potential through precision clock}\label{sec5}
The ratio of electronic transition frequencies between two species is influenced by the nuclear charge radius (field shift) and nuclear mass (mass shift) \cite{Dzuba:2024pri}. For heavy nuclei, the field shift term is predominant, while for light nuclei, the mass shift term plays a more significant role. Any variation in transition frequencies can be attributed to changes in the gravitational potential. In this context, the variation is examined under the influence of a constant DM density and a scalar mediated long-range potential. 

Therefore \cite{Dzuba:2024pri}, 
\begin{equation}
\frac{1}{k}\frac{\delta(\nu_a/\nu_b)}{\nu_a/\nu_b}= \delta\Big(\frac{2\pi}{3}\rho_0 r^2\Big)=\frac{4\pi}{3}\rho_0 r\delta r,  \label{clock1}
\end{equation}
where $k$ is a constant factor depends on the atomic species $a$ and $b$ and $\delta r$ denotes the variation of the distance between the source and the gyroscope due to the ellipticity of the orbit. 

Similarly, for the scalar mediated long range potential, we can write
\begin{equation}
  \frac{1}{k}\frac{\delta(\nu_a/\nu_b)}{\nu_a/\nu_b}= \frac{g^2qQ}{4\pi M_{\mathrm{gy}}r^2}(1+m_\varphi r)e^{-m_\varphi r}\delta r.
  \label{clock2}
\end{equation}
Terrestrial and space--based precision clocks can analyze the variations in the ratio of atomic or nuclear transition frequencies caused by the ellipticity of the planet or satellite orbit can provide constraints on DM or potential new long-range forces \cite{Dzuba:2024pri}.
\section{Sagnac effect in dark matter filled background}\label{sec6}
The Sagnac effect occurs when the propagation of light is influenced by the rotation of a system, such as in ring interferometers. In this setup, light from a source (satellite) is split into two beams: one traveling in the direction of the  gravitational Source's rotation and the other moving in the opposite direction. Due to the rotation, the beam moving with the gravitational source takes longer to return, while the one traveling against the rotation returns more quickly. When these two beams recombine at the source, they produce an interference pattern. The time difference between the beams depends on the angular velocity of the gyroscope or the spacecraft and the area enclosed by the light's path.

We compute the Sagnac time delay in the presence of the DM-induced Schwarzschild spacetime, as described in Eq. \ref{eq:1}. We consider the planar motion of the test object with a constant orbital radius $r=R$ around the source, restricted to the plane defined by $\theta=\pi/2$. In the rotating reference frame, we apply the transformation $\phi\rightarrow \phi-\Omega t$, where $\Omega$ represents angular velocity of the rotating reference frame. Incorporating this angular displacement, the metric becomes
\begin{equation}
ds^2=-\Big(1-\frac{2M}{R}+\frac{4\pi}{3}\rho_0 R^2-R^2\Omega^2\Big)dt^2+R^2d\phi^2-2R^2\Omega dt d\phi.
\label{sag1}
\end{equation}
The observer measures the invariant proper time or the Sagnac time influenced by the DM background as
\begin{equation}
\delta\tau_{\mathrm{DM}}=-4\pi\frac{g_{t\phi}}{\sqrt{g_{tt}}}=-\frac{4\pi R^2\Omega}{\sqrt{1-\frac{2M}{R}+\frac{4\pi}{3}\rho_0R^2-R^2\Omega^2}}. 
\label{sag2}
\end{equation}
Therefore, the relative deviation of the Sagnac time due to constant DM density compared to the Schwarzschild background as
\begin{equation}
\delta_{\mathrm{Sagnac}}=\Big|\frac{\delta\tau_{\mathrm{DM}}-\delta\tau_{\mathrm{Sch}}}{\delta\tau_{\mathrm{Sch}}}\Big|=\Bigg|1-\frac{\sqrt{1-\frac{2M}{R}-R^2\Omega^2}}{\sqrt{1-\frac{2M}{R}+\frac{4\pi}{3}\rho_0R^2-R^2\Omega^2}}\Bigg|,   
\label{deviation}
\end{equation}
where in the limit, $\rho_0=0$, $\delta\tau_{\mathrm{DM}}$ reduces to $\delta\tau_{\mathrm{Sch}}$.
\section{Results}\label{sec8}
In this section, we obtain limits on the DM density and the scalar-electron coupling by analyzing several precision measurements. These include the geodetic precession of gyroscope spin, searches for new forces and non-gravitational potentials using precision clocks and Sagnac time measurements.
  
The strength of these limits is significantly influenced by the mass of the gravitating source object that generates geodetic and frame-dragging precessions, as well as the distance from this gravitating source. Larger distances from the source mass results in more stringent limits on DM density and electrophilic scalar coupling. 

\subsection{Limits on DM density and scalar-electron coupling from the geodetic effect}
\begin{table}[h]
\centering
\begin{tabular}{ |l|c|c|c|c|c|c|c|}
 \hline
 \multicolumn{5}{|c|}{Geodetic effect} \\
 \hline
Orbit (Source)\hspace{0.01cm} & Mass (Source) (GeV)\hspace{0.01cm}&Orb. radius (AU)\hspace{0.01cm}&$(\rho_0/\bar{\rho})$\hspace{0.01cm}& $g$\\
 \hline
GP-B (Earth) & $3.35\times 10^{51}$& $4.69\times 10^{-5}$ &$1.05\times10^{22}$ & $1.96\times 10^{-20}$ \\
 \hline
Earth (Sun) & $1.12\times 10^{57}$ & $1$ & $6.80\times 10^9$ & $8.51\times 10^{-23}$ \\
\hline
Neptune (Sun) & $1.12\times 10^{57}$ & $30$  & $1.24\times 10^9$ & $5.99\times 10^{-21}$ \\
\hline
\end{tabular}
\caption{\label{table1}Limits of DM density and scalar-electron coupling from geodetic effect}
\end{table}
The geodetic drift rate due to the source object is modified by the effects of DM density and the scalar mediated long-range force between the source and the gyroscope. In TABLE \ref{table1}, we present the limits on DM overdensity derived from measurements of the GP-B gyroscope in Earth's orbit \cite{Everitt:2011hp}, as well as from geodetic precession in both Earth and Neptune orbits, influenced by the presence of the Sun \cite{Giorgini1996,PashkevichVershkov+2022+77+109}. We find that the DM density near the GP-B orbital radius is constrained by $\eta =\rho_0/\bar{\rho} \lesssim 1.05 \times 10^{22}$. For Earth's orbit, at approximately 1 AU from the Sun, the sensitivity on the DM density is tighter, with $\eta \lesssim 6.80 \times 10^9$. In Neptune's orbit, at a distance of around 30 AU, the DM density is constrained further to $\eta \lesssim 1.24 \times 10^9$, where we consider that the geodetic drift rates for the Earth and the Neptune due to the Sun are measured with an accuracy of $\mathcal{O}(10^{-7})$ \cite{PashkevichVershkov+2022+77+109}. The DM density constraint becomes more stringent at larger orbital distances.

Additionally, we derive bounds on the scalar coupling, $g$, based on the GP-B motion around Earth. For this case, the upper limit on the scalar coupling is $g \lesssim 1.96 \times 10^{-20}$, which holds for a scalar mass $m_\varphi \lesssim 2.82 \times 10^{-14}$ eV. For Earth's orbit, the limit is $g \lesssim 8.51 \times 10^{-23}$ for a scalar mass $m_\varphi \lesssim 1.32 \times 10^{-18}$ eV. Similarly, for Neptune's orbit, we find $g \lesssim 5.99 \times 10^{-21}$ for $m_\varphi \lesssim 4.40 \times 10^{-20}$ eV. The mass of the scalar is constrained by the inverse of the orbital radius.
\subsection{Limits on DM density and scalar-electron coupling from the search for a new force}
\begin{table}[h]
\centering
\begin{tabular}{ |l|c|c|c|c|c|c|c|}
 \hline
 \multicolumn{5}{|c|}{New force} \\
 \hline
Orbit (Source)\hspace{0.01cm} & Mass (Source)\hspace{0.01cm} (GeV)&Orb. radius\hspace{0.01cm} (AU)&$\rho_0/\bar{\rho}$\hspace{0.01cm}& $g$\\
 \hline
 GP-B (Earth) &$3.35\times 10^{51}$ &$4.69\times 10^{-5}$ & $5.76\times 10^{20}$  & $2.90\times 10^{-21}$ \\
 \hline
Earth (Sun) & $1.12\times 10^{57}$ & $1$ & $1.99\times10^{13}$ & $2.90\times 10^{-21}$ \\
\hline
Neptune (Sun) & $1.12\times 10^{57}$ & $30$ & $7.39\times 10^{8}$& $2.90\times 10^{-21}$\\
\hline
\end{tabular}
\caption{\label{table2}Limits of DM density and scalar-electron coupling from the search for a new force}
\end{table}
The non-gravitational potential generated by DM and ultralight scalar could manifest as a novel force. This force can be probed in fifth force experiments and through tests of the equivalence principle. We summarize the limits on DM overdensity and scalar coupling in TABLE \ref{table2}.

To constrain DM overdensity, we use Eq. \ref{force2}, which relates the force induced by DM. Specifically, we evaluate the ratio of the DM-induced force acting between the detector and the source object to the standard gravitational force. This ratio provides a measure of the relative strength of the non-gravitational interaction, offering insight into the presence and we obtain
\begin{equation}
 \frac{F_{\mathrm{DM}}}{F_G}=\frac{4\pi\rho_0 r^3}{3M_S},
 \label{res1}
\end{equation}
where $M_S$ represents the mass of the source object. Assuming that the contribution of the DM-induced force is within the measurement uncertainty of the source mass, estimated to be approximately $10^{-4}$ for both the Earth and the Sun \cite{uncertainty}, we derive bounds on DM overdensity. For the GP-B orbit, the limit is $\eta\lesssim 5.76\times 10^{20}$, while for Earth's orbit, it tightens to $\eta\lesssim 1.99\times 10^{13}$. In the case of Neptune's orbit, the limit becomes $\eta\lesssim 7.39\times 10^{8}$. The sensitivities are applicable at their respective orbital radii. 

To obtain bounds on the scalar-electron coupling, we use Eq. \ref{force4}. The ratio of the scalar mediated force to the gravitational force is given as 
\begin{equation}
\frac{F_\varphi}{F_G}=\frac{g^2 Qq}{4\pi GM_S M_t}, 
\label{res2}
\end{equation}
where $Q$  and $q$ represent the number of electrons in the source object (Earth or Sun) and the test object, respectively, with masses $M_S$ and $M_t$ and we reinsert $G$. The ratio of the scalar mediated force to the gravitational force is independent of the orbital radius. Given that the measurement uncertainty in the masses of the Earth and Sun is comparable, we derive a uniform limit on the scalar coupling, $g\lesssim 2.90\times 10^{-21}$, applicable in all three scenarios.
\subsection{Limits on DM density and scalar-electron coupling from the search for non-gravitational potential}
\begin{table}[h]
\centering
\begin{tabular}{ |l|c|c|c|c|c|c|c|}
 \hline
 \multicolumn{5}{|c|}{Search for non-gravitational potential} \\
 \hline
Orbit (Source)\hspace{0.01cm} & Mass (Source)\hspace{0.01cm} (GeV)&Orb. radius (AU)\hspace{0.01cm}&$\rho_0/\bar{\rho}$\hspace{0.01cm}& $g$\\
 \hline
 GP-B (Earth) & $3.35\times 10^{51}$ &$4.69\times 10^{-5}$ & $3.24\times 10^{17}$ & $8.16\times 10^{-23}$ \\
 \hline
Earth (Sun) & $1.12\times 10^{57}$ & $1$ & $5.89\times 10^7$ & $7.09\times 10^{-24}$ \\
\hline
Neptune (Sun) & $1.12\times 10^{57}$ & $30$ &  $1.39\times 10^5$ & $3.78\times 10^{-23}$\\
\hline
\end{tabular}
\caption{\label{table3}Limits of DM density and scalar-electron coupling from the search for non-gravitational potential}
\end{table}
The gravitational potential experienced by a satellite orbiting a stellar object can deviate from the standard Newtonian potential due to the influence of DM density and a potential long-range scalar mediated force. A high-precision clock placed aboard the satellite or a terrestrial clock can detect such non-gravitational contributions by monitoring variations in the satellite or planetary orbit, particularly changes in distance caused by orbital ellipticity.

In TABLE \ref{table3}, limits are obtained on the DM overdensity and the scalar-electron coupling strength, considering various celestial bodies like the Earth and the Sun as the source objects. Precision space and ground based clocks in different orbits—such as the GP-B satellite's orbit around Earth, Earth and Neptune around the Sun are capable of detecting deviations from expected potentials. Considering Neptune's orbit is crucial for obtaining tighter limits on the DM overdensity, as it is the most distant known planet in the Solar System. Its vast orbital distance allows for more significant deviations from standard gravitational effects, enhancing the sensitivity of constraints derived from its dynamics. These deviations could reveal the presence of non-gravitational interactions. 

Using Eq. \ref{clock1}, we derive limits on the DM density by analyzing variations in the distance between the source and the detector caused by orbital ellipticity. For the GP-B satellite, with an orbital radius of $4.69\times 10^{-5}~\mathrm{AU}$ and an eccentricity of $0.0014$, the bound on the DM overdensity relative to local DM density is found to be $\eta\lesssim 3.24\times 10^{17}$ for $r\sim 4.69\times 10^{-5}~\mathrm{AU}$. Precision clocks in Earth's orbit around the Sun further refine this constraint. For Earth's orbital radius around the Sun of $1~\mathrm{AU}$ and an eccentricity of $0.017$, the bound tightens to $\eta\lesssim 5.89\times 10^{7}$.

In the case of Neptune orbit, assuming an orbital radius of $\sim 30~\mathrm{AU}$, and an eccentricity of $0.008$, the limit becomes $\eta\lesssim 1.39\times 10^{5}$ for $r\sim 30~\mathrm{AU}$. These results highlight that increasing the orbital radius yields stricter limits on DM density, leveraging the sensitivity of precision clocks to non-gravitational forces in a variety of orbital configurations.

To determine bounds on the scalar-electron coupling, we utilize Eq. \ref{clock2}. For the GP-B orbit, the coupling is constrained to $g\lesssim 8.16\times 10^{-23}$, valid for $m_\varphi\lesssim 2.82\times 10^{-14}~\mathrm{eV}$. Similarly, for Earth's orbit around the Sun, the coupling is limited to $g\lesssim 7.09\times 10^{-24}$, applicable when $m_\varphi\lesssim 1.32\times 10^{-18}~\mathrm{eV}$. For the Neptune orbit, the limit becomes $g\lesssim 3.78\times 10^{-23}$, assuming $m_\varphi\lesssim 4.40\times 10^{-20}~\mathrm{eV}$. These constraints highlight the sensitivity of orbital dynamics to potential deviations from standard gravity due to scalar-electron interactions.

We derive constraints on $\eta$ and $g$ by comparing the variation in the non-gravitational potential, induced by DM and the long-range force, with the current sensitivity of nuclear clocks $(10^{-19})$ \cite{Campbell:2012zzb}. As seen in TABLE \ref{table3}, the bounds on DM overdensity become more stringent with increasing orbital radius. While the constraints obtained for the GP-B orbit are less stringent, they remain significant because they apply to small orbital radii, a regime not extensively explored in previous studies.

In contrast, the constraints on the scalar coupling $g$ behaves differently with the orbital radius as compared to the DM density case. The strength of these bounds is influenced by the number of charged particles present in the source object, with larger charge content in the source resulting in tighter constraints. This dependence highlights the interplay between distance and source properties in deriving limits on new physics.
\subsection{Limits on DM density from Sagnac effect}
\begin{table}[h]
\centering
\begin{tabular}{ |l|c|c|c|c|c|}
 \hline
 \multicolumn{4}{|c|}{Sagnac effect} \\
 \hline
Orbit (Source)\hspace{0.01cm} & Mass (Source) (GeV)\hspace{0.01cm}&Orb. radius (AU)\hspace{0.01cm}&$(\rho_0/\bar{\rho})$\hspace{0.01cm}\\
 \hline
GP-B (Earth) & $3.35\times 10^{51}$ & $4.69\times 10^{-5}$ & $1.82\times 10^{15}$ \\\hline
Earth (Sun) & $1.12\times 10^{57}$ & $1$  & $4.01\times 10^{6}$ \\
 \hline
Neptune (Sun) & $1.12\times 10^{57}$ & $30$  & $4.45\times 10^{3}$ \\
 \hline
\end{tabular}
\caption{\label{table4}Limits of DM density from Sagnac effect}
\end{table}
We obtain limits on the DM overdensity by measuring the Sagnac time using the precision clock aboard a satellite in orbits around Earth and Neptune, as well as the GP-B orbit. To calculate the relative deviation in Sagnac time due to a constant DM density, we use Eqs. \ref{eq:a2} and \ref{deviation}, which compare the deviation from the Schwarzschild background. Therefore, we obtain
\begin{equation}
\delta_{\mathrm{Sagnac}}\simeq \frac{2\pi}{3}\rho_0 R^2,
\label{res3}
\end{equation}
which implies that a larger orbital radius leads to a stronger constraint on the DM density. In TABLE \ref{table4}, we present the constraints on the DM overdensity obtained from measurements of the precision nuclear clock onboard satellites in different orbits. For the GP-B orbit, the constraint is $\eta\lesssim 1.82\times 10^{15}$, for Earth's orbit $\eta\lesssim 4.01\times 10^{6}$, and for Neptune's orbit $\eta\lesssim 4.45\times 10^{3}$.

\begin{figure}[htp]
    \centering
    \includegraphics[width=12cm]{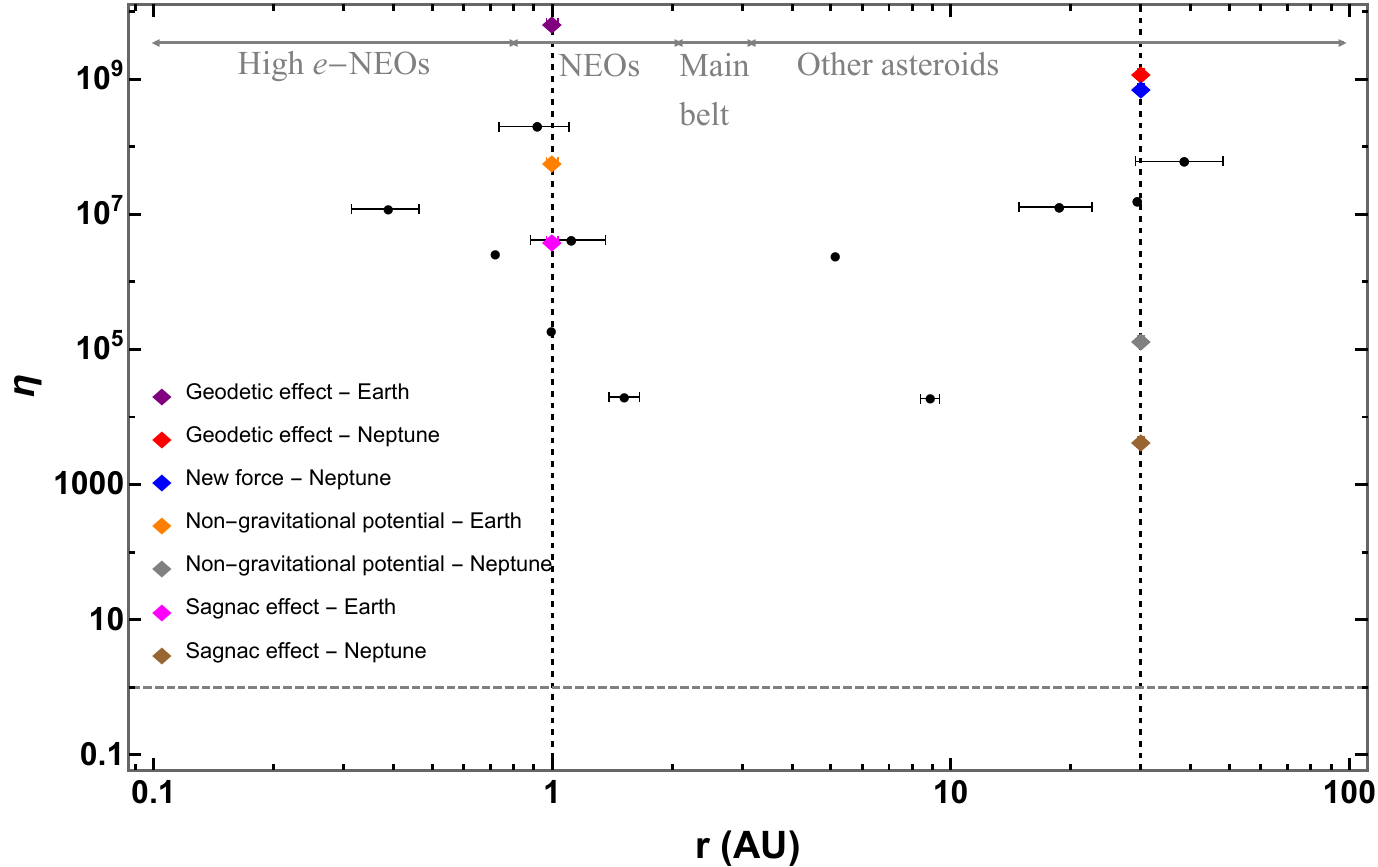}
    \caption{Limits on DM overdensity as a function of orbital radius. The black circles represent previously established constraints on DM overdensity \cite{Anderson:1995dw,Gron:1995rn,Pitjev:2013sfa,Tsai:2022jnv}, while the colored diamonds correspond to the results obtained in this study. The length of the horizontal bar denotes the distance between the perihelion and aphelion positions.}
    \label{one}
\end{figure}
In FIG. \ref{one}, we present limits on DM overdensity as a function of orbital radius, derived from analyses of the geodetic effect, new forces, non-gravitational potentials, and the Sagnac effect. Each point (circles and diamonds) represents the DM overdensity at a specific orbital radius, providing distinct bounds on $\eta$. 

The gray solid arrows indicate the locations of Near-Earth Objects (NEOs), including highly eccentric objects, main belt asteroids, and other bodies. Tracking these objects can be used to probe DM density. The gray horizontal dashed line corresponds to $\eta=1$, dividing the plot into two regions: the area above the dashed line represents DM overdensity $(\eta>1)$, while the region below it corresponds to DM underdensity $(\eta<1)$.

In the orbit of Neptune, a satellite equipped with a precision nuclear clock can impose significant constraints on the DM overdensity. The most stringent limit, $\eta\lesssim 4.45\times 10^{3}$ arises from the Sagnac effect (brown diamond). Additional sensitivities include $\eta\lesssim 1.39\times 10^{5}$ from the search for non-gravitational potential (gray diamond), $\eta\lesssim 7.39\times 10^{8}$ from the search for a new force (blue diamond), and $\eta\lesssim 1.24\times 10^{9}$ from the search for the geodetic effect (red diamond).

A satellite equipped with a precision nuclear clock in Earth's orbit can place significant limits on the DM. The tightest limit, $\eta\lesssim 4.01\times 10^{6}$ is derived from the Sagnac effect (magenta diamond). Additional limits include $\eta\lesssim 5.89\times 10^{7}$ from the search for a non-gravitational potential (orange diamond) and $\eta\lesssim 6.80\times 10^{9}$ from the geodetic effect (purple diamond). 

\begin{figure}[htp]
    \centering
    \includegraphics[width=12cm]{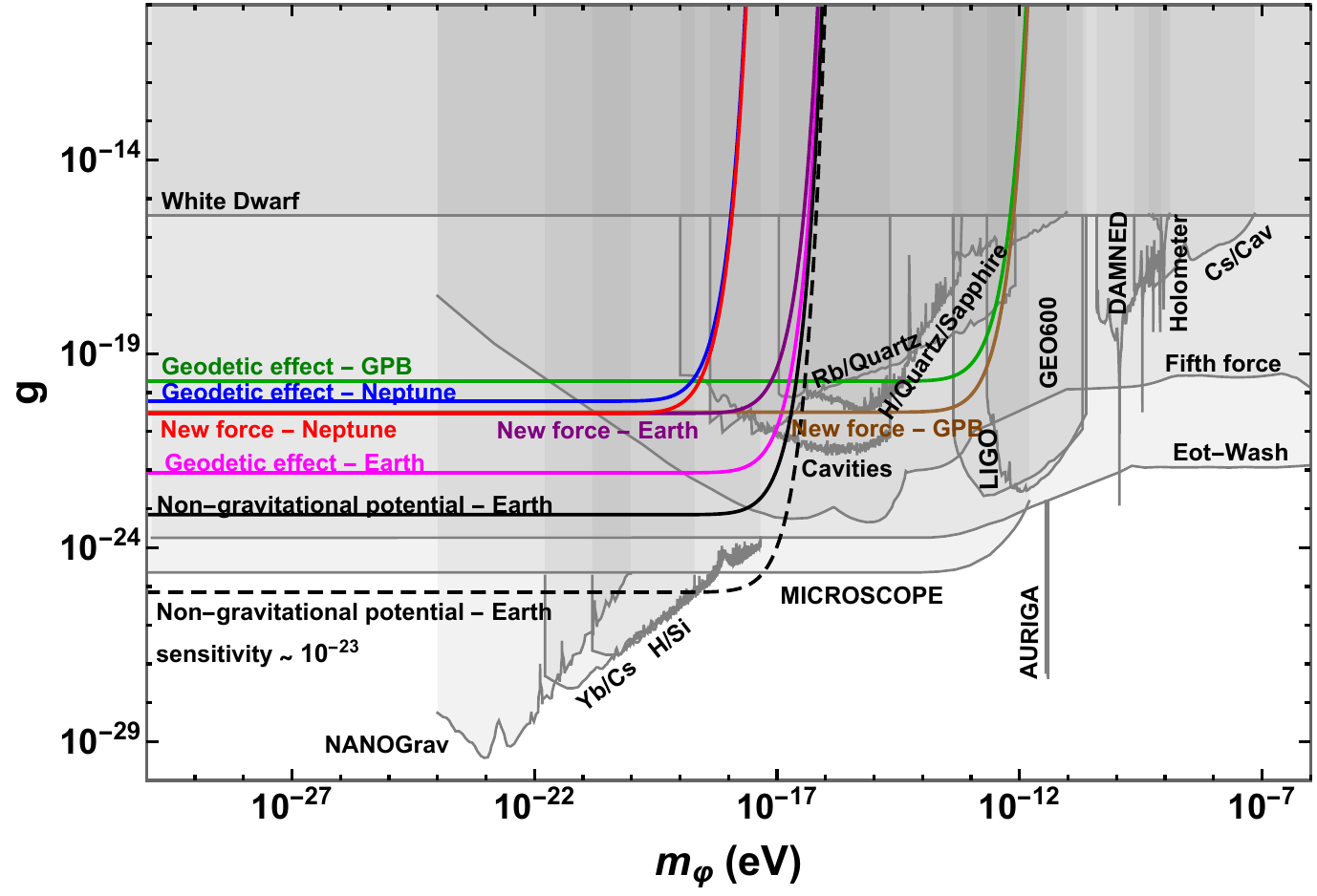}
    \caption{Limits on the scalar-electron coupling of the long-range force as a function of the ultralight scalar mass. The gray shaded regions represent previously established limits \cite{Baryakhtar:2017ngi,Bustamante:2018mzu,Wise:2018rnb}, while the colored lines indicate the results obtained in this study.}
    \label{two}
\end{figure}

In FIG. \ref{two}, we present limits on the scalar-electron coupling of the long-range force as a function of the ultralight scalar mass. The interaction arises from the presence of electrons within the source object and the gyroscope or spacecraft, allowing us to probe scalar mediated Yukawa forces.

The analysis of GP-B data, which tracked Earth's motion, places constraints on the scalar coupling, yielding $g\lesssim 1.96\times 10^{-20}$ for $m_\varphi\lesssim 2.82\times 10^{-14}~\mathrm{eV}$ (green line). Investigating the geodetic effect with a precision of $\mathcal{O}(10^{-7})$ for Neptune and Earth’s motion around the Sun, using precision gyroscopes onboard satellites, provides projected bounds of $g\lesssim 5.99\times 10^{-21}$ (blue line) and $g\lesssim 8.51\times 10^{-23}$ (magenta line) for $m_\varphi\lesssim 4.40\times 10^{-20}~\mathrm{eV}$ and $m_\varphi\lesssim 1.32\times 10^{-18}~\mathrm{eV}$ respectively.

Similarly, the search for a new force gives comparable constraints on $g$ with $g\lesssim 2.90\times 10^{-21}$ for different scalar masses, determined by the orbital radii (GP-B: brown line; Earth: purple line; Neptune: red line). Among these, the search for a non-gravitational potential with nuclear clocks in Earth's orbit sets the most stringent limit, $g\lesssim 7.09\times 10^{-24}$ (black line). 

This result is highly competitive with existing fifth-force and Eöt-Wash experiment constraints. Notably, space-based clocks with sensitivity at the level of $\mathcal{O}(10^{-23})$ could potentially surpass the current limits set by the MICROSCOPE experiment \cite{Berge:2017ovy}.

The scalar mass is inversely constrained by the orbital radius of the test system. Future space-based precision measurements hold the potential to reach sensitivity levels sufficient to place constraints on $g$ that exceed the current limits.

In Appendix \ref{sec7}, we also leverage the GINGER experiment \cite{Altucci:2022rxr,DiVirgilio:2023nrc} to constrain the DM density within the Earth. The GINGER experiment operates based on the Sagnac effect and is designed to measure the frame-dragging effect on spacetime caused by the Earth's rotation. This Earth-based experiment calculates the difference in proper time between two counter-propagating beams traveling in a rotating frame. we utilize data from the GINGER experiment to estimate the DM overdensity within the Earth. However, with the current sensitivity of the experiment, the resulting constraints are not particularly optimistic. If the experimental sensitivity improves to the level of nuclear clock precision, around $\mathcal{O}(10^{-19})$, it would allow us to derive bounds on the DM overdensity comparable in magnitude to those obtained from the geodetic drift measurements of the GP-B experiment.
\section{Complementary search for cosmic neutrino density}\label{addneu}
Non-relativistic cosmic neutrinos can gravitationally cluster, leading to the formation of a cosmic neutrino overdensity. The SM predicts a total number density of cosmic neutrinos of approximately $\bar{n}_\nu \simeq 336/\mathrm{cm^3}$. Given that the energy of these neutrinos is extremely low (ranging from $10^{-4}~\mathrm{eV}$ to $10^{-6}~\mathrm{eV}$), the massive neutrinos in the SM are non-relativistic and can cluster gravitationally, leading to the creation of overdensities in space. The cosmic neutrino overdensity can be scaled in a manner analogous to the DM overdensity as
\begin{equation}
\xi=\frac{\rho_\nu}{\bar{\rho}_\nu}=1.2\times 10^{7}\Big(\frac{\rho_0}{\bar{\rho}}\Big)\Big(\frac{0.1~\mathrm{eV}}{m_\nu}\Big),    
\end{equation}
where $m_\nu$ denotes the neutrino mass. 
\begin{table}[h]
\centering
\begin{tabular}{ |l|c|c|c|c|c|c|c|}
 \hline
 \multicolumn{3}{|c|}{Cosmic neutrino overdensity $(\xi)$} \\
 \hline
Orbit (Source)\hspace{0.01cm} & Non-gravitational potential\hspace{0.01cm}&Sagnac effect\\
 \hline
Earth (Sun) & $7.07\times 10^{14}$ & $4.81\times 10^{13}$ \\
\hline
Neptune (Sun) & $1.67\times 10^{12}$ & $5.34\times 10^{10}$ \\
\hline
\end{tabular}
\caption{\label{tableneutrino}Limits on cosmic neutrino overdensity from the search for non-gravitational potential and Sagnac effect}
\end{table}

In Table \ref{tableneutrino}, we present constraints on cosmic neutrino overdensity derived from the search for non-gravitational potentials and the Sagnac effect in both Earth's and Neptune's orbits. Only the scenarios yielding the strongest limits on cosmic neutrino overdensity are considered. From the search for non-gravitational potential, we find $\xi \lesssim 7.07 \times 10^{14}$ at approximately 1 AU and $\xi \lesssim 1.67 \times 10^{12}$ at about 30 AU. Additionally, we obtain $\xi \lesssim 4.81 \times 10^{13}$ for Earth's orbit and $\xi \lesssim 5.34 \times 10^{10}$ for Neptune's orbit from the Sagnac time measurements. The limits are derived for $m_\nu\sim0.1~\mathrm{eV}$. Some of these bounds are competitive with, or even stronger than, the global upper limit on cosmic neutrino density set by the KATRIN experiment ($\xi\lesssim 10^{11}$) \cite{KATRIN:2022kkv} and local neutrino overdensity from cosmic rays ($\xi\lesssim 10^{13}$) \cite{Ciscar-Monsalvatje:2024tvm}. It is important to note that the Pauli exclusion principle imposes a limit on cosmic neutrino overdensity, with an upper bound of $\xi \lesssim 10^{6}$, although this limit can be relaxed by introducing new neutrino interactions \cite{Bauer:2022lri}.
\section{Conclusions and discussions}\label{sec9}
In this paper, we obtain limits on overdensities of DM and cosmic neutrinos, and the scalar-electron coupling using measurements of the geodetic effect, new force searches, non-gravitational potentials, and the Sagnac effect, at various planetary and gyroscope orbits within the solar system.

A constant DM density within the orbit of the GP-B gyroscope around the Earth, or within the orbits of Earth and Neptune around the Sun, can influence the geodetic drift rate, which is measured within the framework of GR. By analyzing existing data and assuming that any DM contribution should lie within the GR measurement uncertainty, we obtain limits on the local DM overdensity at specific orbital radii.

Additionally, a constant DM overdensity within a given orbital radius induces non-gravitational potentials and forces, which can be constrained through terrestrial and space--based precision clock experiments. We also derive limits on DM overdensity from Sagnac time measurements using precision clocks placed in orbits. These DM overdensity constraints can be rescaled to infer bounds on cosmic neutrino overdensity. Notably, the constraints strengthen with increasing orbital radius.

Key results include:
\begin{itemize}
\item From the Sagnac effect at the GP-B orbit, we limit the DM overdensity to $\eta\lesssim 1.82\times 10^{15}$, marking one of the first reported bounds at such a small length scale $(7027.4~\mathrm{km})$.

\item At Earth’s orbital radius $(1~\mathrm{AU})$, the strongest limit is derived from the Sagnac effect, with $\eta\lesssim 4.01\times 10^{6}$ which is competitive with the existing bound.

\item At Neptune’s orbital radius $(30~\mathrm{AU})$, we obtain $\eta\lesssim 4.45\times 10^{3}$, which is four orders of magnitude stronger than existing limits. 
\end{itemize}

We also place limits on cosmic neutrino overdensity through non-gravitational potential measurements and the Sagnac effect. The latter, conducted at Neptune’s orbit, limits the cosmic neutrino overdensity to $\xi\lesssim 5.34\times 10^{10}$, improving upon previous limits from cosmic ray studies and KATRIN by factors of three and one, respectively. 

The presence of electrons in the gravitating source and the gyroscope or spacecraft introduces a long-range Yukawa-type scalar mediated force between the source and the test object, in addition to gravity. This force influences the motion of the gyroscope or spacecraft and can be investigated through geodetic effects, as well as searches for non-gravitational potentials and new forces.

The strongest limit on the scalar-electron coupling, $g\lesssim 7.09\times 10^{-24}$, is derived from the search for a non-gravitational potential using precision clock measurements for $m_\varphi\lesssim 1.32\times 10^{-18}~\mathrm{eV}$. This constraint is comparable to existing fifth-force and Eöt-Wash limits. Future advancements in precision measurements up to a sensitivity of $\mathcal{O}(10^{-23})$ could further strengthen these bounds on the coupling, potentially exceeding the limits set by MICROSCOPE. 

The limits on DM and neutrino overdensities are influenced by the assumed DM and neutrino density profile. In our analysis, we adopt the simplest scenario of a constant DM density. However, alternative DM density profiles, such as NFW, Burkert, or Einasto, would yield different bounds on the DM overdensity. These models introduce additional parameters, making the results dependent on those parameters.

The key advantage of our approach with a constant DM density is its simplicity—there are no additional parameters involved. Consequently, the limits on DM overdensity in our analysis are independent of any specific model assumptions, providing a more straightforward and robust baseline for comparison.

The sensitivities on DM density and the scalar-electron coupling derived from the search for non-gravitational potential are influenced by variations in the orbital distance between the source and the test object. Orbits with higher eccentricity yield more stringent limits on both the DM density and the scalar coupling. Additionally, the limits on DM density become increasingly tighter with larger orbital radii such as Trans Neptunian Objects (TNOs).

As a result, tracking objects with highly eccentric orbits and greater orbital radii can provide even stronger bounds on DM density. Future space-based missions and precision clock experiments in space are expected to significantly enhance these limits on both density and coupling parameters. These studies not only serve to test GR on larger scales but also offer a means to explore DM and various particle physics models, particularly in cases where direct detection constraints are limited or unavailable.

Constraints on DM overdensity and the scalar coupling have been derived from GP-B data through measurements of the geodetic effect. Similarly, bounds can be obtained using data from other missions such as LARES \cite{Capozziello:2014mea,Ciufolini:2023czv}, LAGEOS \cite{Iorio:2002rm,Ciufolini:2004rq,Iorio:2008vm}, GRACE \cite{doi:10.1126/science.1099192}, Lunar Laser Ranging (LLR) \cite{Bertotti:1987zz,Williams:1995nq}, STE-QUEST \cite{Tsai:2021lly,STE-QUEST:2022eww}, and GRACE-FO \cite{Abich:2019cci}. Additionally, ground-based experiments like Very Long Baseline Interferometry (VLBI) \cite{Klopotek2020,SCHUH201268} and Global Navigation Satellite Systems (GNSS) \cite{Cuadrat-Grzybowski:2024uph,Bertrand:2023zkl,Heki2023} offer complementary opportunities to derive constraints on DM density and scalar coupling. 

Beyond the Solar System, the geodetic effect can also be investigated in systems such as double pulsar binaries or the S2 star, which orbits Sagittarius A* with a semi-major axis of approximately $1000~\mathrm{AU}$. For the S2 star, we estimate that the constraint on $\eta$ could reach $\lesssim \mathcal{O}(10^2)$, assuming measurement precision comparable to that of nuclear clock. Deploying quantum sensors, such as accelerometers, on spacecraft in the orbit of the S2 star \cite{Iorio:2017fck} or a double pulsar binary \cite{Breton:2008xy} could significantly enhance navigational precision. 

Future space-based clocks in planetary orbits offer promising avenues for probing new physics, providing stringent constraints on various BSM couplings \cite{Tsai:2021lly}. Advanced space-based quantum clocks, such as NASA's Deep Space Atomic Clock (DSAC) \cite{Burt2021}, have demonstrated the potential for deployment in space. These clocks not only enable autonomous spacecraft navigation, relativistic geodesy, and secure quantum communication \cite{doi:10.1126/science.aan3211} but also serve as tools to test GR and explore DM.

Recent advancements include NASA's Parker Solar Probe reaching an orbit as close as $0.05~\mathrm{AU}$ to the Sun, the MESSENGER mission tracking Mercury's orbit \cite{Cavanaugh2007}, and the JUNO spacecraft currently collecting data around Jupiter \cite{Li:2022wix,Singh:2024nou}. The upcoming Neptune Odyssey mission \cite{Rymer_2021} will complement these efforts by placing tighter constraints on DM overdensity and scalar-electron coupling, while also testing GR and the SM of particle physics.

The Space Quantum Clock Networks (SQCN), a proposed system connecting space-based and Earth-based quantum clocks, promises enhanced precision and reduced calibration uncertainties \cite{Wcislo:2018ojh,Derevianko:2021wgw,Shen:2021scn}. Such networks can refine measurements of DM density and GR effects. Additionally, the Sagnac effect—arising as a phase shift between atoms in different internal states transported along closed paths in opposite directions—provides a unique mechanism for testing fundamental physics without free propagation \cite{PhysRevLett.115.163001}. Novel techniques have been proposed that leverage the coupling between photon spin and gyroscope rotation, enabling the measurement of gravitomagnetic fields around celestial bodies \cite{Fedderke:2024ncj, Mashhoon:2024wvp}.

In this work, we focus on the impact of new physics on geodetic precession. The influence of new physics on frame-dragging effects is another intriguing avenue of study, which we plan to address in a future publication. The study in this paper underscores the synergy between astrophysical observations, laboratory experiments, and precision metrology in probing new physics beyond the SM and GR.
\section*{Acknowledgement} S.R.A. would like to thank COST Actions CA23130 (BridgeQG) and CA22113 (THEORY-CHALLENGES) supported by COST (European Cooperation in Science and Technology). G.L. and T.K.P. would like to thank COST Actions 
CA21106 (COSMIC WISPers), and CA23130 (BridgeQG) supported by COST (European Cooperation in Science and Technology).

\appendix
\section{Christoffel symbols for Schwarzschild spacetime in presence of constant DM background}\label{app1}
The Christoffel symbol is defined as
\begin{equation}
 \Gamma^{\alpha}_{\beta\gamma}=\frac{1}{2}g^{\alpha\sigma}\Big(\frac{\partial g_{\sigma\beta}}{\partial x^\gamma}+\frac{\partial g_{\sigma\gamma}}{\partial x^\beta}-\frac{\partial g_{\beta\gamma}}{\partial x^\sigma}\Big),   
\end{equation}
where the non-zero Christoffel symbols for the metric Eq. \ref{eq:1} in planar motion are
\begin{equation}
\begin{split}
\Gamma^0_{10}=\Big(1-\frac{2M}{r}+\frac{4\pi}{3}\rho_0 r^2\Big)^{-1}\Big(\frac{M}{r^2}+\frac{4\pi}{3}\rho_0 r\Big),~~~\Gamma^1_{11}=-\Big(1-\frac{2M}{r}-\frac{8\pi}{3}\rho_0r^2\Big)^{-1}\Big(\frac{M}{r^2}-\frac{8\pi\rho_0 r}{3}\Big),\\
\Gamma^1_{00}=\Big(1-\frac{2M}{r}-\frac{8\pi}{3}\rho_0r^2\Big)\Big(\frac{M}{r^2}+\frac{4\pi}{3}\rho_0 r\Big),~~~\Gamma^1_{33}=-r\Big(1-\frac{2M}{r}-\frac{8\pi}{3}\rho_0r^2\Big),~~~\Gamma^3_{13}=\frac{1}{r},\\
\Gamma^1_{22}=-r\Big(1-\frac{2M}{r}-\frac{8\pi}{3}\rho_0r^2\Big),~~~\Gamma^2_{12}=\frac{1}{r}.
\end{split}
\label{eq:2}
\end{equation}

\section{Parallel transport of gyroscope spin in presence of long-range Yukawa potential}\label{app2}
Let us consider the spin of a gyroscope $S^\mu =\left(S^{(0)},S^{(1)},S^{(2)},S^{(3)}\right)$ and in the rest frame, $u^\mu =\left(u^0,\bold{0}\right)$, $S^\mu = \left(0,\bold{S}\right)$. Therefore $S\cdot u=0$, which implies at any frame
\begin{equation}
    g_{\mu\nu}u^\mu S^\nu=0.
    \label{ap2a}
\end{equation}
Differentiating Eq. \ref{ap2a} with respect to the proper time $\tau$, we obtain
\begin{equation}
    \frac{d}{d\tau}\left(g_{\mu\nu}u^\mu S^\nu\right) = g_{\mu\nu,\alpha}\,u^\alpha u^\mu S^\nu+g_{\mu\nu} \frac{d u^\mu}{d\tau}S^\nu+g_{\mu\nu}u^\mu\,\frac{dS^\nu}{d\tau} =0,
    \label{appendix_1stepEqSpin}
\end{equation}
\noindent where we use $\frac{d g_{\mu\nu}}{d\tau} =  \frac{d x^\alpha}{d\tau}\frac{\partial  g_{\mu\nu}}{\partial x^\alpha} \equiv g_{\mu\nu,\alpha}\,u^\alpha$. 

If we consider any four potential $A_\mu$ coupled with the four current, then Eq. \ref{geodetic+interaction} generalizes to 
\begin{equation}
   \ddot{x}^\alpha+\Gamma^\alpha_{\mu\nu}\dot{x}^\mu\dot{x}^\nu=\frac{g\,q}{M_{gy}}g^{\alpha\mu}(\partial_\mu A_\nu-\partial_\nu A_\mu)\,\dot{x}^\nu.
    \label{geodetic+interactionnew}
\end{equation}
Therefore, using Eq. \ref{geodetic+interactionnew} in Eq. \ref{appendix_1stepEqSpin}, we obtain
\begin{equation}
    g_{\mu\nu,\alpha}\,u^\alpha u^\mu S^\nu+g_{\mu\nu}S^\nu\left[-\Gamma^\mu_{\alpha\beta}u^\alpha u^\beta +\frac{g\,q}{M_{gy}}g^{\mu\alpha}\left(\partial_\alpha A_\beta-\partial_\beta A_\alpha\right)u^\beta \right]+g_{\mu\nu}u^\mu\,\frac{dS^\nu}{d\tau} =0,
\end{equation}
which implies 
\begin{equation}
     g_{\mu\nu,\alpha}\,u^\alpha u^\mu S^\nu-\Gamma^\mu_{\alpha\beta}u^\alpha u^\beta S_\mu + \frac{g\,q}{M_{gy}}\delta^\alpha_\nu\left(\partial_\alpha A_\beta-\partial_\beta A_\alpha\right)u^\beta S^\nu +g_{\mu\nu}u^\mu\,\frac{dS^\nu}{d\tau} =0 .
    \label{appendix_2stepEqSpin}
\end{equation}
\noindent Note that,
\begin{align*}
    -\Gamma^\mu_{\alpha\beta}u^\alpha u^\beta S_\mu +   g_{\mu\nu,\alpha}\,u^\alpha u^\mu S^\nu =& -\frac{1}{2} g^{\mu\lambda}\left(g_{\lambda\alpha,\beta}+g_{\lambda\beta,\alpha}-g_{\alpha\beta,\lambda}\right) u^\alpha u^\beta S_\mu + g_{\mu\nu,\alpha}\,u^\alpha u^\mu S^\nu \\
    =& -\frac{1}{2} S^\lambda\left(g_{\lambda\alpha,\beta}+g_{\lambda\beta,\alpha}-g_{\alpha\beta,\lambda}\right) u^\alpha u^\beta + \underbrace{g_{\mu\nu,\alpha}\,u^\alpha u^\mu S^\nu}_%
    {\mu\rightarrow\beta,\,\nu\rightarrow\lambda} \\
    =&-\frac{1}{2} S^\lambda\left(g_{\lambda\alpha,\beta}-g_{\lambda\beta,\alpha}-g_{\alpha\beta,\lambda}\right) u^\alpha u^\beta \\
    =& \,\frac{1}{2} g^{\mu\lambda}\left(g_{\lambda\alpha,\beta}-g_{\lambda\beta,\alpha}-g_{\alpha\beta,\lambda}\right) u^\alpha u^\beta S_\mu \\
    =& \,\frac{1}{2} g^{\mu\lambda}\left(g_{\lambda\alpha,\beta}-g_{\lambda\beta,\alpha}-g_{\alpha\beta,\lambda}\right) u^\alpha g^{\gamma\beta}u_\gamma S_\mu \\
    =& \,\frac{1}{2} g^{\gamma\beta}\left(g_{\lambda\alpha,\beta}-g_{\lambda\beta,\alpha}-g_{\alpha\beta,\lambda}\right)g^{\mu\lambda} u_\gamma u^\alpha S_\mu \\
    =& \Gamma^\gamma_{\lambda\alpha} u_\gamma u^\alpha S^\lambda.
\end{align*}

\noindent Replacing the result obtained in Eq. \ref{appendix_1stepEqSpin} ($\gamma\rightarrow\mu$, $\lambda\rightarrow\alpha$, $\alpha\rightarrow\beta$), we get
\begin{equation*}
    g_{\mu\nu} u^\mu\,\frac{dS^\nu}{d\tau}+\Gamma^\mu_{\alpha\beta}S^\alpha u^\beta u_\mu = -\frac{g\,q}{M_{gy}}\delta^\alpha_\nu\left(\partial_\alpha A_\beta-\partial_\beta A_\alpha\right) u^\beta S^\nu
\end{equation*}
\begin{equation*}
    \Longleftrightarrow u_\mu \,\frac{dS^\mu}{d\tau} +\Gamma^\mu_{\alpha\beta}S^\alpha u^\beta u_\mu =  -\frac{g\,q}{M_{gy}}g^{\mu\beta}\left(\partial_\alpha A_\beta-\partial_\beta A_\alpha\right) S^\alpha u_\mu
\end{equation*}
\begin{equation*}
    \Longrightarrow \frac{dS^\mu}{d\tau} +\Gamma^\mu_{\alpha\beta}u^\alpha S^\beta = \frac{g\,q}{M_{gy}}g^{\mu\alpha}\left(\partial_\alpha A_\beta-\partial_\beta A_\alpha\right) S^\beta.
\end{equation*}
\noindent For the scalar field, $A_\mu$ will be replaced by $\varphi$ and that corresponds to Eq. \ref{spin_eq_Yukawa}.
\section{Dark Matter Effects on Beat Frequency Measurements in the GINGER Experiment}\label{sec7}
GINGER (Gyroscopes IN GEneral Relativity) is an advanced ring laser gyroscope array designed to measure Earth's frame-dragging effect with exceptional precision using the Sagnac effect. By detecting the time difference between counter-propagating light beams with an accuracy of $10^{-4}$, GINGER enables stringent tests of gravitational theories. As the first experiment optimized to observe Earth's gravitomagnetic effect, it is uniquely suited to probe spacetime distortions caused by Earth's rotation.

The experiment's beat frequency measures GR effects tied to Earth's rotation. A ring laser converts time differences into frequency differences. Since, the emission is continuous, each beam of light are standing waves with wavelengths, given as $c\tau_+=P=N\lambda_+$ and $c\tau_-=P=N\lambda_-$, with $N$ the integer and $P$ the length of the loop. Note, these modes can have different $N$, but the higher accuracy has been reached with equal $N$. Therefore, the proper time difference of two standing waves can be written as
\begin{equation}
\delta\tau=N(\lambda_+-\lambda_-)=N\frac{f_--f_+}{f^2}=P\lambda\delta f.
\label{gen13}
\end{equation}
Similarly, we can write the ring-laser equation as 
\begin{equation}
\delta f=\frac{4A}{\lambda P}\mathbf{u}_n\cdot \mathbf{\Omega},
\label{gen14}
\end{equation}
where the total rotation rate at leading order is obtained as 
\begin{equation}
    \boldsymbol{\Omega} = -\boldsymbol{\Omega}_\oplus+2\left(\frac{M}{r}+\frac{4\pi}{3}\rho_0\,\frac{r^3_0}{r}\right)\sin\vartheta\,\Omega_\oplus \,\boldsymbol{u}_\vartheta+\frac{1}{r^3}\left[\boldsymbol{J}-3\left( \boldsymbol{J}\cdot\boldsymbol{u}_r\right)\boldsymbol{u}_r\right],
    \label{final_omega}
\end{equation}
where $\boldsymbol{J} =\boldsymbol{J}_\oplus+\boldsymbol{J}_{DM}$ and $\boldsymbol{J}_i=I_i\,\boldsymbol{\Omega}_\oplus$, with $i=\oplus,DM$.
It is useful to use orthonormal spherical basis $\{\mathbf{u}_r, \mathbf{u}_\vartheta, \mathbf{u}_\varphi\}$ in the Earth-fixed inertial frame so that $\vartheta=\pi/2$ corresponds to the equatorial plane. If $\alpha$ denotes the angle between the radial direction $\mathbf{u}_r$ and the normal vector $\mathbf{u}_n=\mathbf{u}_r\cos\alpha+\mathbf{u}_\vartheta\sin\alpha$, then we can write the expression for the beat frequency in presence of DM within the Earth and obtain
\begin{equation}
\begin{split}
\delta f=\frac{4A}{\lambda P}\Big[\Omega_\oplus\cos(\vartheta+\alpha)-\Big(\frac{2M}{r}+\frac{8\pi}{3}\rho_0\frac{r^3_0}{r}\Big)\Omega_\oplus\sin\vartheta\sin\alpha+\frac{(I_\oplus+I_\mathrm{DM})}{r^3}\Omega_\oplus(2\cos\vartheta\cos\alpha+\\
\sin\vartheta\sin\alpha)\Big],    
\label{gen17}
\end{split}
\end{equation}
where $\Omega_\oplus$ denotes the angular velocity of the Earth and $I_\mathrm{DM}=(2/5)\cdot(4/3)\pi r^3_0\rho_0\cdot r^2_0$. Here, $r_0$ represents the radius of Earth's core, where DM may exist due to its purely gravitational interaction. Additionally, the measurement of Earth's core mass involves significant uncertainties.

We derive constraints on DM overdensity using the GINGER experiment, designed to measure Earth's frame-dragging effect with an accuracy of $10^{-4}$ \cite{Capozziello:2021goa}. The time difference between counter-propagating laser beams, $\delta\tau$, includes both GR and DM contributions, while $\delta\tau_\mathrm{GR}$ accounts for only the GR effect. Therefore, 
\begin{equation}
\Big|\frac{\delta\tau}{\delta\tau_\mathrm{GR}}-1\Big|<10^{-4}, 
\label{gin1}
\end{equation}
where we constrain the DM overdensity to $\eta\lesssim 1.79\times 10^{31}$, assuming Earth's core radius $r_0\sim 2000~\mathrm{km}$ and total radius $R\sim 6400~\mathrm{km}$. While current GINGER sensitivity provides a pessimistic bound, achieving nuclear clock precision $\mathcal{O}(10^{-19})$, could tighten the bound to $\eta\lesssim 1.79\times 10^{16}$.

\bibliographystyle{utphys}
\bibliography{reference_spin}
\end{document}